\documentclass[conference]{IEEEtran}

\usepackage[ruled]{algorithm2e}
\usepackage{stmaryrd}
\usepackage{graphicx}
\usepackage{color}
\usepackage{comment}
\usepackage{url}
\usepackage{wrapfig}
\usepackage{subfigure}
\usepackage{tabularx}
\usepackage{algpseudocode}

\newcommand{\superscript}[1]{\ensuremath{^{\textrm{#1}}}}

\def\wu{\superscript{*}}
\def\wg{\superscript{\dag}}

\begin{document}

\title{Adaptive 360 VR Video Streaming\\ based on MPEG-DASH SRD}

\author{\IEEEauthorblockN{Mohammad Hosseini\wu\wg, Viswanathan Swaminathan\wg}
\IEEEauthorblockA{
%  \sharedaffiliation
  \begin{tabular}{ccc}
    \wu University of Illinois at Urbana-Champaign (UIUC) & &\wg Adobe Research, San Jose, USA \\
  \end{tabular}
  ~\\
Email: shossen2@illinois.edu, vishy@adobe.com}}

\maketitle

\begin{abstract}
We demonstrate an adaptive bandwidth-efficient 360 VR video streaming system based on MPEG-DASH SRD. We extend MPEG-DASH SRD to the 3D space of 360 VR videos, and showcase a dynamic view-aware adaptation technique to tackle the high bandwidth demands of streaming 360 VR videos to wireless VR headsets. We spatially partition the underlying 3D mesh into multiple 3D sub-meshes, and construct an efficient 3D geometry mesh called \textit{hexaface sphere} to optimally represent tiled 360 VR videos in the 3D space. We then spatially divide the 360 videos into multiple tiles while encoding and packaging, use MPEG-DASH SRD to describe the spatial relationship of tiles in the 3D space, and prioritize the tiles in the Field of View (FoV) for view-aware adaptation. Our initial evaluation results show that we can save up to 72\% of the required bandwidth on 360 VR video streaming with minor negative quality impacts compared to the baseline scenario when no adaptations is applied.

\end{abstract}
\IEEEpeerreviewmaketitle
%\vspace{-.5cm}

\section{Introduction}
%With the advances in VR technologies and the corresponding growth in wireless VR headsets, there has been significant interest towards 360-degree VR video applications in the recent years. 
360 VR videos are immersive spherical videos, mapped into a 3D geometry where the user can look around during playback using a VR head-mounted display (HMD). 
%That gives viewer a sense of depth in every direction. 
Unfortunately 360 VR videos are extremely bandwidth intensive especially with the 4K video resolution being widely viewed as a functional minimum resolution for current HMDs, and 8K or higher desired. Therefore, a major challenge is how to efficiently transmit these bulky 360 VR videos to bandwidth-constrained wireless VR HMDs at acceptable quality levels given their high bitrate requirements.

In this work, we are motivated by the 360 VR video applications with 8K and 12K resolutions and the data rate issues that such rich multimedia system have. We extend MPEG-DASH SRD  towards the 3D VR environment, and showcase how to utilize a semantic link between the users' viewport, spatial partitioning, and stream prioritization using a divide and conquer approach. Once 360 videos are captured, we spatially \textit{divide} them into multiple video tiles while encoding and packaging, which are efficiently textured on the underlying 3D geometry mesh that we construct called \textit{hexaface sphere}. We then use MPEG-DASH SRD to describe the spatial relationship of tiles in the 3D space, and develop a prioritized view-aware approach to \textit{conquer} the intense bandwidth requirements. We showcase our demo using a real-platform wireless HMD and multiple 360 VR video sequences, and show that our adaptations significantly reduces the total bandwidth required to deliver a high quality immersive experience. Our approach can further increase the overall 360 VR video quality at a given bandwidth, virtually allowing 8K and higher VR video resolutions.

\begin{figure}[!t]
\centering
\includegraphics[width=.5\columnwidth]{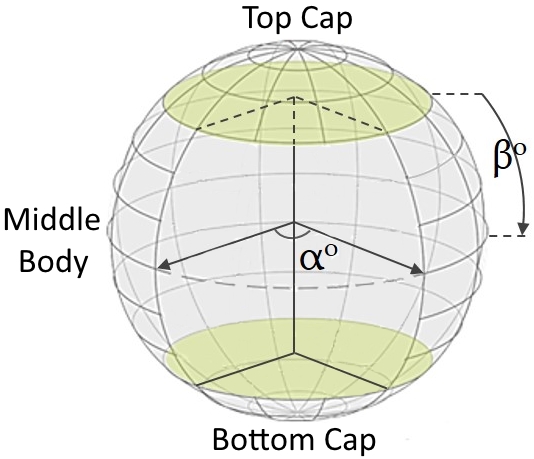}~~\includegraphics[width=.45\columnwidth]{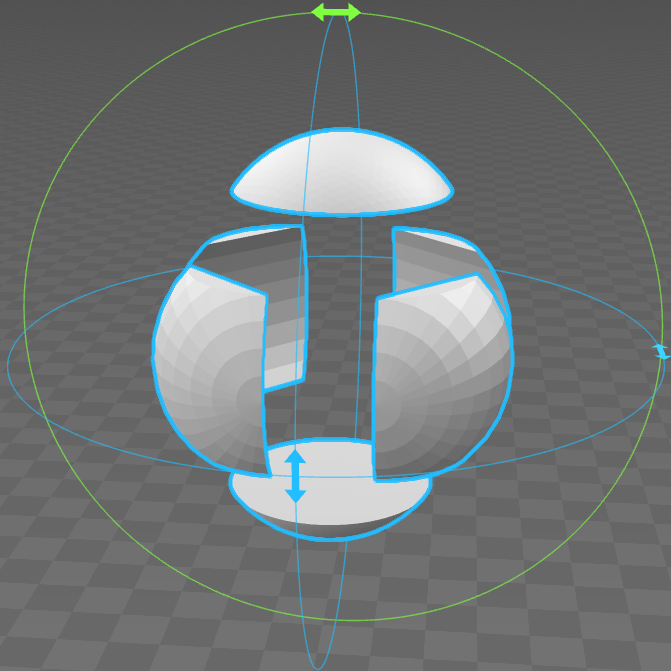}
\caption{Visual overview of a generated hexaface sphere.}
\label{hexaface}
\vspace{-.3cm}
\end{figure}

\begin{figure}[!t]
\centering
\includegraphics[width=.42\textwidth]{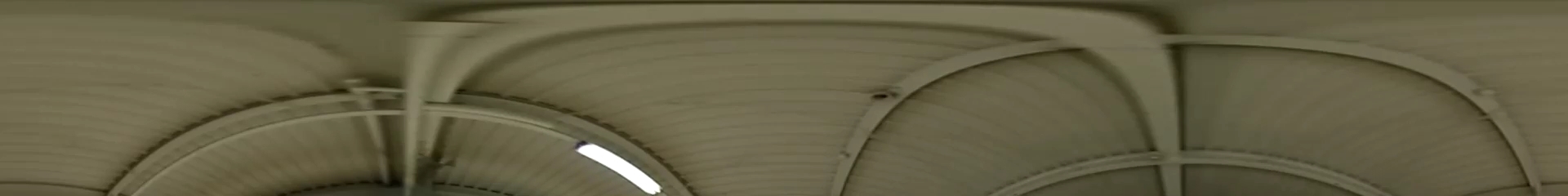}\\
~\\
\includegraphics[width=.1\textwidth]{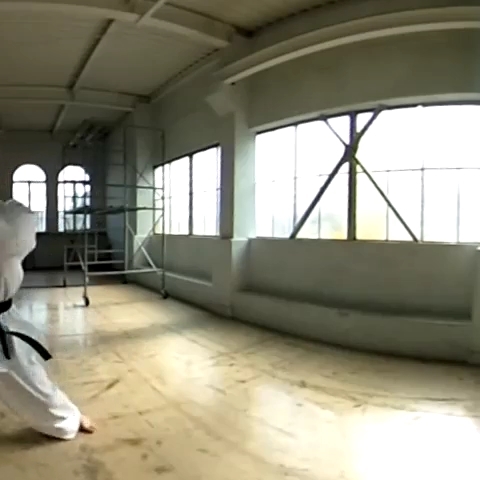}~~~\includegraphics[width=.1\textwidth]{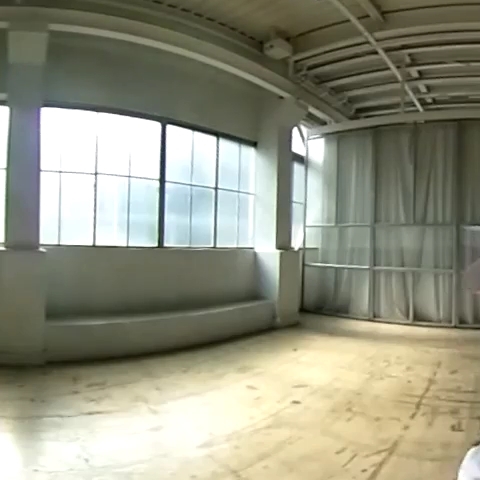}~~~\includegraphics[width=.1\textwidth]{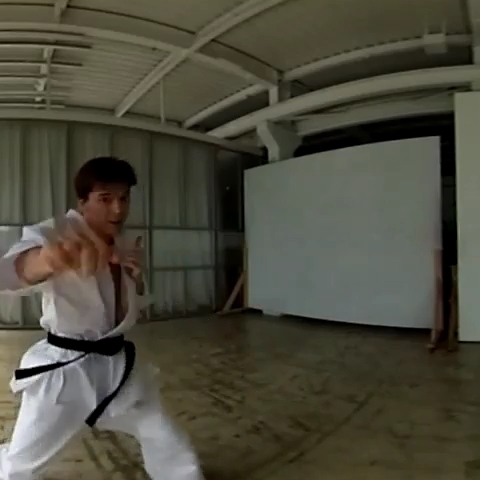}~~~\includegraphics[width=.1\textwidth]{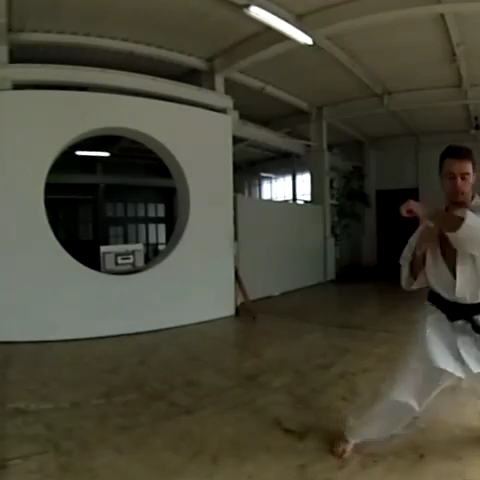}\\
~\\
\includegraphics[width=.42\textwidth]{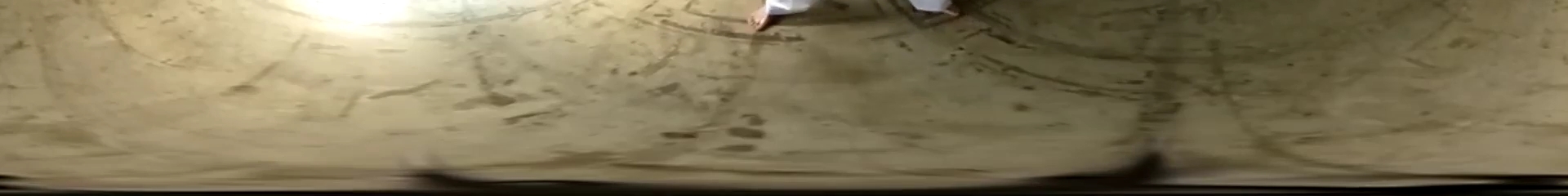}\\
\caption{Various tiles of an example 360 video (\textit{Karate}) according to the six 3D meshes of our \textit{hexaface sphere} 3D geometry.}
\label{karate}
\vspace{-.5cm}
\end{figure}

\section{Methodology}
%360 VR videos are created by mapping a raw 360 video as a 3D texture onto a 3D geometry mesh, often a sphere, with the user at the center of that geometry. 
While watching a 360 degree video on a VR HMD, a user views only a small portion of the 360 degrees. That small portion is equivalent to a specific confined region on the underlying 3D spherical mesh which is \textit{spatially} related to the corresponding portion of the raw content. For example, the Samsung Gear VR HMD offers a 96-degree FoV, meaning it can only cover a quarter of a whole 360-degree-wide content horizontally.

To decrease the bandwidth requirements of 360 VR videos, we use a prioritized view-aware technique, and stream tiles inside the viewport at highest resolution, at or near the native resolution of the HMD. To achieve that, our approach consists of two parts. First, we spatially partition the 360 video into multiple tiles. We extend the features of MPEG-DASH SRD towards the 3D space, and define a reference space for each video tile, corresponding to the rectangular region encompassing the entire raw 360-degree video. Second, we partition the underlying 3D geometry into multiple segments, each representing a sub-mesh of the original 3D mesh with a unique identifier, in a two-step process. In the first step, using the concepts of \textit{slices} and \textit{stacks} used in spherical 3D reconstruction, we split the sphere programatically into 3 major parts, including the top cap, the middle body, and the bottom cap. The middle body covers $2\beta^o$ degrees, given the vertical FoV settings of the VR HMD. In the second step, we further split the middle body into four sub meshes, each covering $\alpha^o$ (=$90$ degrees in this case) of the entire 360-degree wide screen given the horizontal FoV settings of the HMD. With this process, our projection will result into a combination of \textit{six} 3D sub-spherical meshes that we call a \textit{hexaface sphere} 3D mesh. Finally, a mapping mechanism is defined for spatial positioning of the tiles on the 3D space, so that each tile be textured on its corresponding 3D mesh segment. Figure \ref{hexaface} illustrates our hexaface sphere 3D geometry. Figure \ref{karate} shows how our tiling process is applied against an example 360 video frame, according to our hexaface sphere geometry.

To enable view awareness, we follow three steps to create valid confines of unit quaternions specifically set for each of the hexaface sphere 3D mesh segments. We first convert Euler angles to a unit quaternion representation for VR device orientation tracking, and calculate an array corresponding to a normalized direction vector from our quaternion. We then combine the values together to compute the confines of 3D segment-specific quaternion representations inside the hexaface sphere. With the confines of each 3D mesh segment being defined, we then identify which 3D segments and the corresponding tiles intersect with a user's viewport and implement our viewport tracking at every frame. With viewport tracking, we then implement our prioritized view-aware adaptation, and dynamically deliver higher bitrate content to the tiles within the user's FoV, and assign lower quality content to the area outside the user's immediate FoV.

\begin{figure}[!t]
\centering
\includegraphics[width=\columnwidth, trim = 50 260 50 250, clip = true]{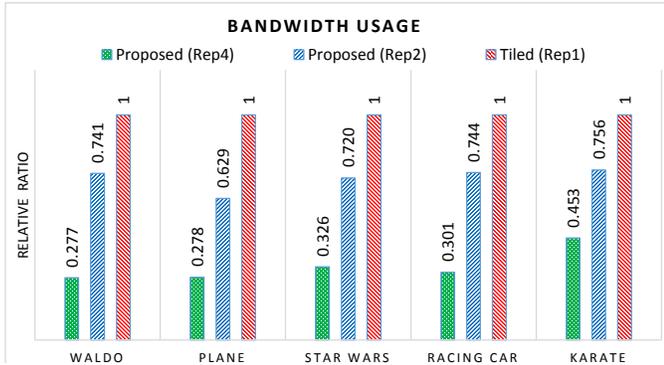}
\caption{A comparison of bandwidth savings of streaming different 360 VR videos, using our adaptations, and tiled-streaming with no adaptation.}
\label{results}
\vspace{-.5cm}
\end{figure}
\section{Evaluation}
To evaluate our work, we used the Samsung Gear VR HMD mounted with the Samsung Galaxy S7 smartphone as our target VR platform. We used the Oculus Mobile SDK 1.0.3 joint with the Android SDK API 24 for development of a 360 VR video streaming application prototype based on MPEG-DASH SRD. We used our developed prototype to apply adaptations and run experiments. Our VR platform provides a total resolution of 2560x1440 (1280x1440 per eye), with maximum frame rate of 60 FPS, and a horizontal FoV of 96 degrees. We set the vertical FoV of our 360 VR video prototype to 90 degrees. We prepared 5 different 360 degree videos with various resolutions publicly available on Youtube as test sequences for the purpose of applying our adaptations.  %We spatially cropped the videos using FFMPEG Zeranoe 64-bit API, and generated 6 different tiles as per requirements of our hexaface sphere geometry. 
We encoded all video tiles into four different representations ($REP_1$ to $REP_4$ with the highest to lowest resolutions), and used MPEG-DASH SRD to describe our tiling.

\begin{figure}[!t]
\centering
\includegraphics[width=.5\textwidth]{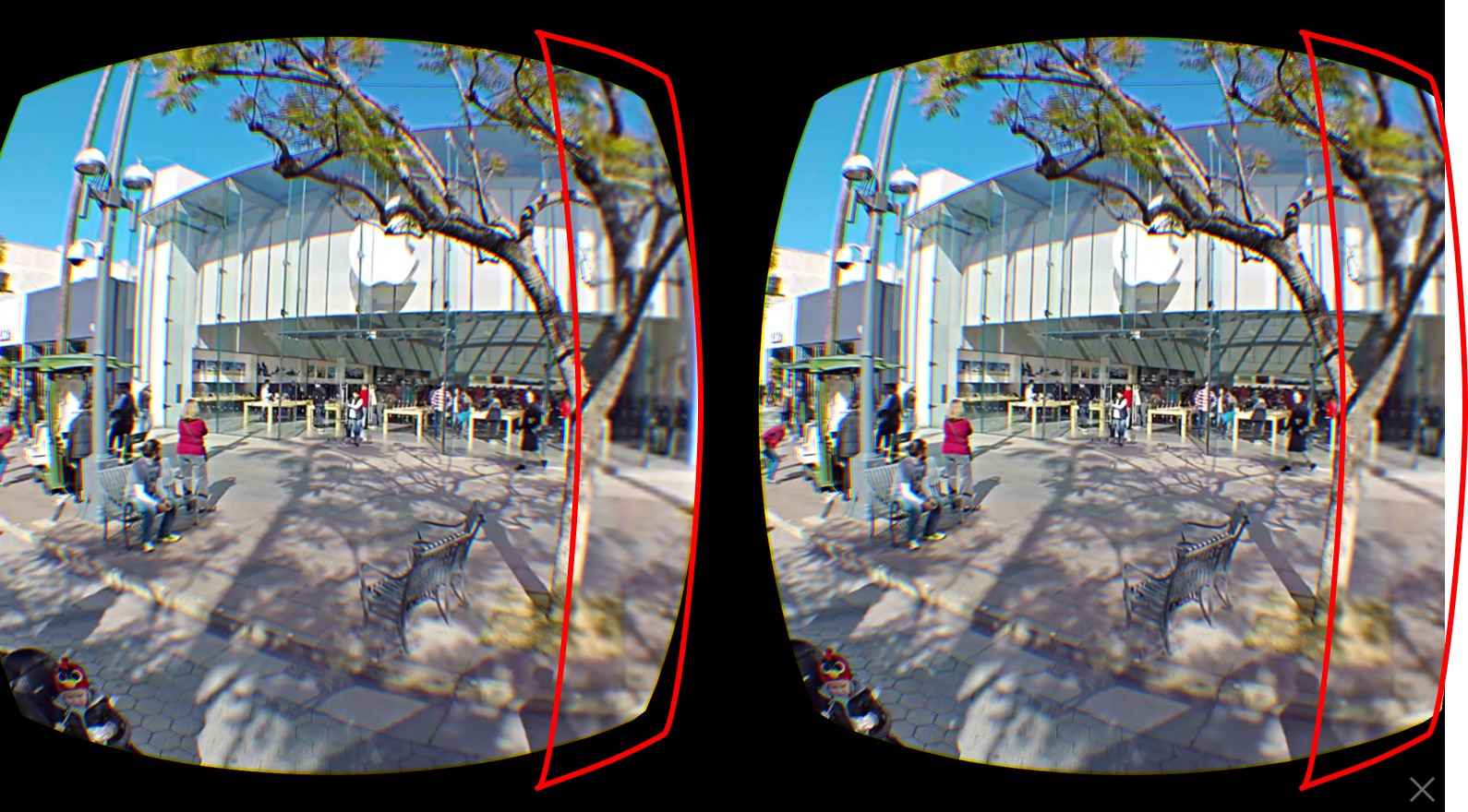}\\
~\\
\includegraphics[width=.5\textwidth]{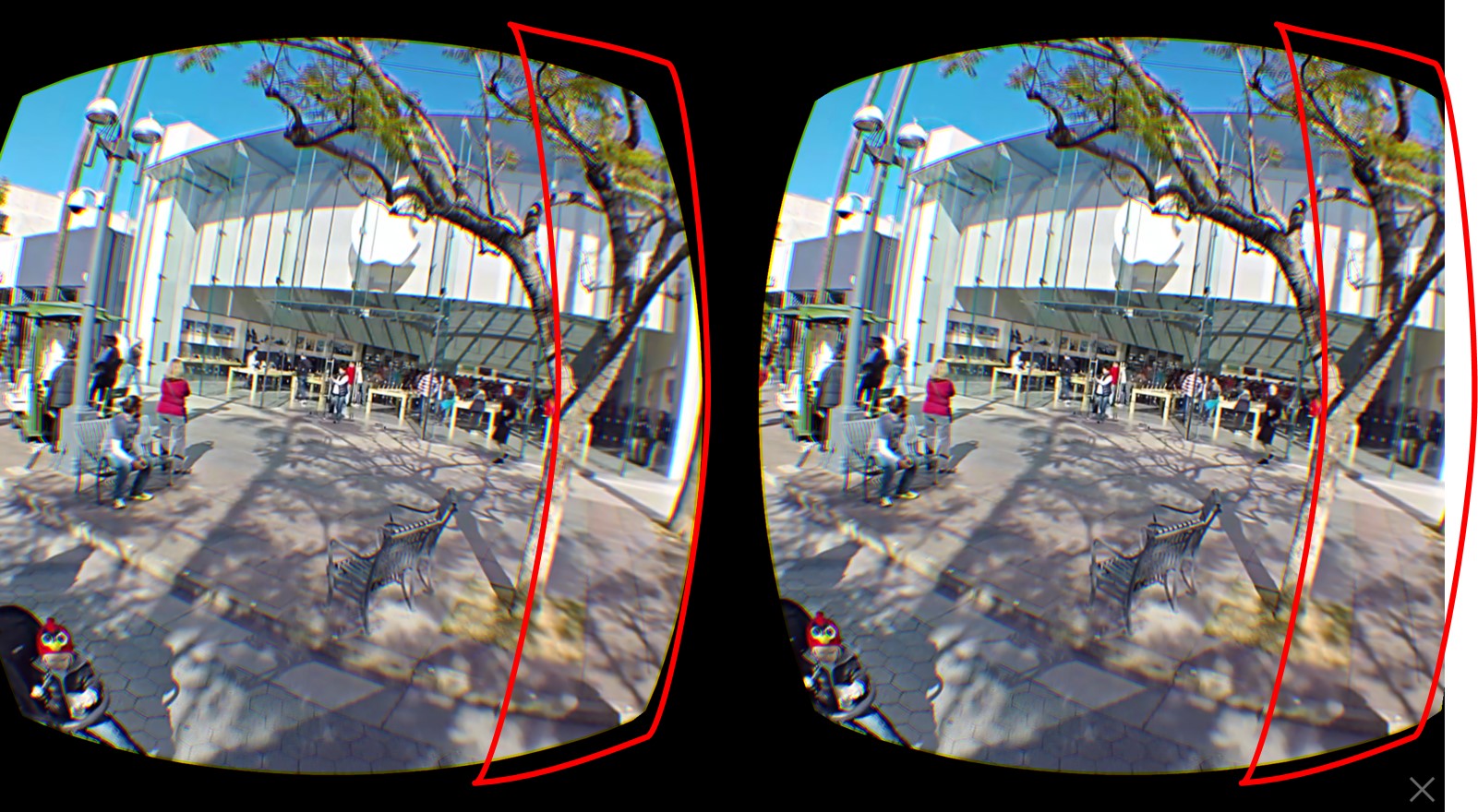}\\
\caption{Visual comparison of a specific frame within a sample 360 VR video with the peripheral tiles having lowest resolution. (Top) $REP_4$ with resolution of 240x480. (Bottom) $REP_2$ with resolution of 720x1440.}
\label{240}
\vspace{-.5cm}
\end{figure}

%We measured the bandwidth usage in terms of average bitrate, when maximum resolution ($REP_4$) is assigned for tiles within immediate FoV, and two different lower resolutions, one with lowest resolution ($REP_1$) and other with second highest resolution ($REP_3$) assigned to the peripheral tiles. 
We compared the relative bandwidth usage when using our adaptations against the baseline case where no adaptation is applied (the 360 VR video is tiled; no view awareness is present, and all tiles are streamed with highest representation, $REP_1$). Figure \ref{results} demonstrates results for only a small subset of our experiments on all of our benchmarks, with ratios normalized. Our results show that our adaptations can significantly save bandwidth usage for up to 72\% compared to the baseline case where our adaptation approach is not employed. Figure \ref{240} shows two sample screenshots of the experiments on \textit{Waldo}. While the highest representation possible ($REP_1$- resolution of 960x1920) is delivered to the main front tile, in Figure \ref{240} (Top) lowest representation is delivered to the peripheral tile on the right ($REP_4$- resolution of 240x480), whereas in Figure \ref{240} (Bottom), the peripheral tile on the right has second highest representation assigned to it ($REP_2$- resolution of 720x1440). The red confines specify the approximate area for the peripheral tile with the lower quality. 

In our demo, we show that even the lowest representation on the peripheral tiles not within immediate viewports results in minor visual changes from a user's perspective, sometimes not even perceptible, while still maintaining the original quality for the main viewport to ensure a satisfactory user experience. Overall, considering the significant bandwidth saving achieved using our adaptations, it is reasonable to believe that many 360 VR video users would accept such minor visual changes given their limited bandwidth. More technical details are presented in our other publication \cite{ism16}.
%The results of our user study will be presented in details in the future work.

\bibliographystyle{IEEEtran}
\bibliography{ref}

\end{document}